\documentclass[12pt,onecolumn]{IEEEtran}
\usepackage{cite}
\usepackage{multirow}
\usepackage{multicol}
\usepackage[linesnumbered,ruled,vlined, lined, commentsnumbered, longend]{algorithm2e} 
\usepackage{amsmath,amssymb,amsfonts}
\usepackage{tablefootnote}
\usepackage[utf8]{inputenc}
\usepackage{graphicx}
\usepackage{hyperref}
\hypersetup{
    colorlinks=true,
    linkcolor=blue,
    filecolor=magenta,      
    urlcolor=cyan,
}
\usepackage{xcolor}

\usepackage{scalerel}
\usepackage{tikz}
\usetikzlibrary{svg.path}
\usepackage{scalerel}[2016/12/29]

\definecolor{orcidlogocol}{HTML}{A6CE39}
\tikzset{
  orcidlogo/.pic={
    \fill[orcidlogocol] svg{M256,128c0,70.7-57.3,128-128,128C57.3,256,0,198.7,0,128C0,57.3,57.3,0,128,0C198.7,0,256,57.3,256,128z};
    \fill[white] svg{M86.3,186.2H70.9V79.1h15.4v48.4V186.2z}
                 svg{M108.9,79.1h41.6c39.6,0,57,28.3,57,53.6c0,27.5-21.5,53.6-56.8,53.6h-41.8V79.1z M124.3,172.4h24.5c34.9,0,42.9-26.5,42.9-39.7c0-21.5-13.7-39.7-43.7-39.7h-23.7V172.4z}
                 svg{M88.7,56.8c0,5.5-4.5,10.1-10.1,10.1c-5.6,0-10.1-4.6-10.1-10.1c0-5.6,4.5-10.1,10.1-10.1C84.2,46.7,88.7,51.3,88.7,56.8z};
  }
}

\newcommand\orcid[1]{\href{https://orcid.org/#1}{\mbox{\scalerel*{
\begin{tikzpicture}[yscale=-1,transform shape]
\pic{orcidlogo};
\end{tikzpicture}
}{|}}}}

\def\BibTeX{{\rm B\kern-.05em{\sc i\kern-.025em b}\kern-.08em
    T\kern-.1667em\lower.7ex\hbox{E}\kern-.125emX}}
    
\begin{document}
\title{MDEAW: A Multimodal Dataset for Emotion Analysis through EDA and PPG signals from wireless wearable low-cost off-the-shelf Devices}

\author{\IEEEauthorblockN{Arijit Nandi\orcid{0000-0003-4238-5183}, Fatos Xhafa\orcid{0000-0001-6569-5497}, Laia Subirats\orcid{0000-0001-8646-5463} and Santi Fort\orcid{0000-0003-2189-6830}}\\
\thanks{Arijit Nandi is with the Department of CS, Universitat Polit\`{e}cnica de Catalunya, 08034 Barcelona, Spain and Eurecat, Centre Tecnol\`{o}gic de Catalunya, 08005 Barcelona, Spain (e-mail: arijit.nandi@eurecat.org ).}
\thanks{ Fatos Xhafa is with the Department of CS, Universitat Polit\`{e}cnica de Catalunya, 08034 Barcelona (e-mail: fatos@cs.upc.edu).}
\thanks{Laia Subirats is with Eurecat, Centre Tecnol\`{o}gic de Catalunya, 08005 Barcelona, Spain and ADaS Lab, Universitat Oberta de Catalunya, 08018 Barcelona, Spain (e-mail: laia.subirats@eurecat.org). }

\thanks{Santi Fort is with Eurecat, Centre Tecnol\`{o}gic de Catalunya, 08005 Barcelona, Spain  (e-mail: santi.fort@eurecat.org). }
}

\maketitle
\begin{abstract}
We present MDEAW, a multimodal database consisting of Electrodermal Activity (EDA) and Photoplethysmography (PPG) signals recorded during the exams for the course taught by the teacher at Eurecat Academy, Sabadell, Barcelona in order to elicit the emotional reactions to the students in a class room scenario. Signals from 10 students were recorded along with the students self-assessment of their affective state after each stimuli, in terms 6 basic emotion states. All the signals were captured using portable, wearable, wireless, low-cost, and off-the-shelf equipment that has the potential to allow the use of affective computing methods in everyday applications. A baseline for student-wise affect recognition using EDA and PPG-based features, as well as their fusion, was established through ReMECS, Fed-ReMECS  and Fed-ReMECS-U. These results indicate the prospects of using low-cost devices for affective state recognition applications. The proposed database will be made publicly available in order to allow researchers to achieve a more thorough evaluation of the suitability of these capturing devices for emotion state recognition applications.
\end{abstract}


\section{Motivation}
The main motivation of conducting this experiment at Eurecat's Augmented Workplace is to see how emotions affect students' learning outcomes. The hypothesis is that positive emotions promote good learning outcomes, while negative emotions could be a reason for bad results. From the common sense perspective, this hypothesis, sounds reasonable. The outcome of this study is to formally prove the hypothesis. Also, to the best of our knowledge there is no suitable dataset available in the emotion research domain which is related to emotional state analysis in the real life E-learning or education context.

\section{Experimental Protocol}
\subsection{Experimental environment}
\begin{figure}[htbp]
    \centering
    \includegraphics[width=0.6\textwidth]{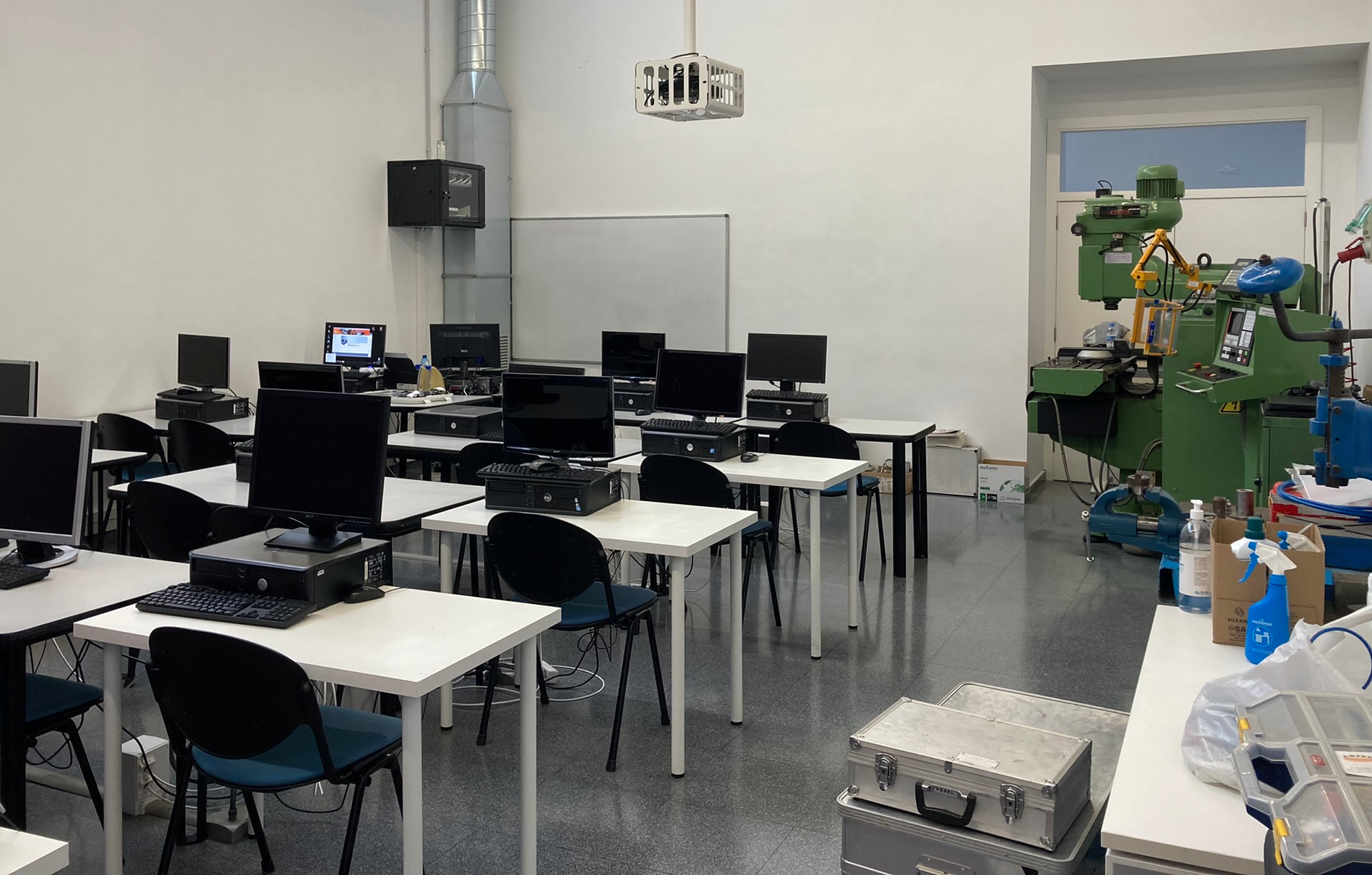}
    \caption{The classroom where the experiment was conducted (Eurecat Academy, Sabadell, Barcelona).}\label{fig:classroom}
\end{figure}

The experiment is conducted at a classroom of Eurecat Academy, Sabadell, Barcelona. The classroom was a regular classroom with fully equipped machineries and other instruments. The picture (Fig.~\ref{fig:classroom}) is the actual view of the classroom where the experiment was conducted.

\subsection{Learning exercise}
The stimuli used in this experiment are the exams for the course taught by the teacher at Eurecat Academy, Sabadell, Barcelona in order to elicit the emotional reactions to the students in a class room scenario and record Electrodermal Activity (EDA) and Photoplethysmography (PPG) data. The exam was designed by the tutor of that course and it was divided into smaller exercises (see Annex~\ref{app:exam} for the examination formulation). 

Each exercise duration was 5 minute slots and there were a total of 12 sets of exercises, so in total the whole exam session was 84 minutes. In between each of the 12 sets we gave 2 minutes gap to for the students to “cool down”.  Also, students were advised that their annotations would not affect their evaluations to avoid any possible bias. In each of the 12 sets, from exercise Set 1 to Set 7; and 11 and 12 had 4 questions in each. Exercise Set 8 to 10 had one subjective question in each. The exercises sets are added in the additional materials for further reference. 

Before starting the exam we instructed the students the whole exam pattern. In which students were asked to give their current emotional states that they had felt after answering each questions. For the emotion states input from students we used 6 basic discrete emotion model~\cite{ekman1992}\footnote{\url{https://www.paulekman.com/universal-emotions/}} developed by Paul Ekman, widely accepted theory of basic emotions and their expressions. These emotion labels are as follows:
\begin{enumerate}
    \item {Sadness}
    \item {Happiness}
    \item {Fear}
    \item {Anger}
    \item {Surprise}
    \item {Disgust}
\end{enumerate}

\subsection{Students}
In the exam there were 10 students. The students were all male. Their age varies from 23 to 57 years. 

\subsection{Data Acquisition}

\begin{figure}[htbp]
    \centering
    \includegraphics[width=0.4\linewidth]{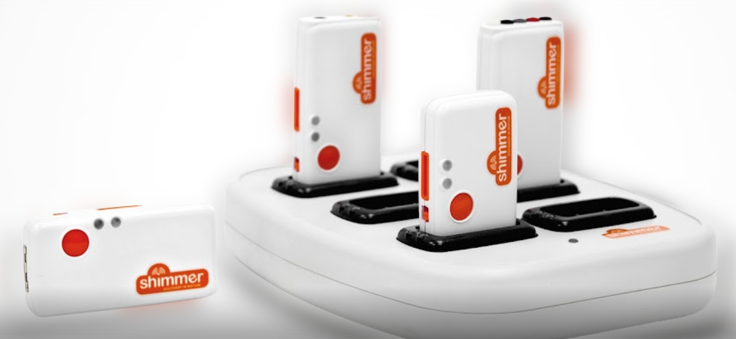}
    \caption{Shimmer Sensors}
    \label{fig:shimmer-sensor}
\end{figure}

The sensors used to record the EDA and PPG data is Consensys Bundle Development kit from Shimmer Sensing. The pictures of the sensor is in Fig~\ref{fig:shimmer-sensor}. 

\begin{figure}[htbp]
    \centering
    \includegraphics[width=0.4\linewidth]{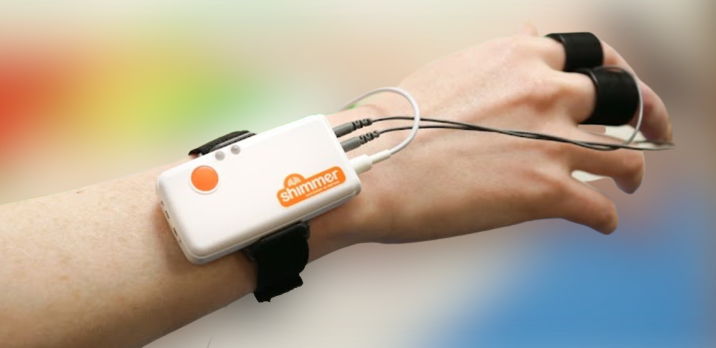}
    \caption{Students wearing Shimmer sensors in the classroom while taking the exams.}
    \label{fig:studenst-with-sensors}
\end{figure}

Before starting the exam we put the sensors on the students shown in the following picture (Fig.~\ref{fig:studenst-with-sensors}) for collecting the EDA and PPG data. The sensors were put on the non active hand of each student (such as if he/she is right handed we put the sensor on the left hand and so on). The whole experiment was carried out in 2 sessions, in each session there were 5 students.

\begin{figure}[htbp]
    \centering
    \includegraphics[width=0.6\linewidth ]{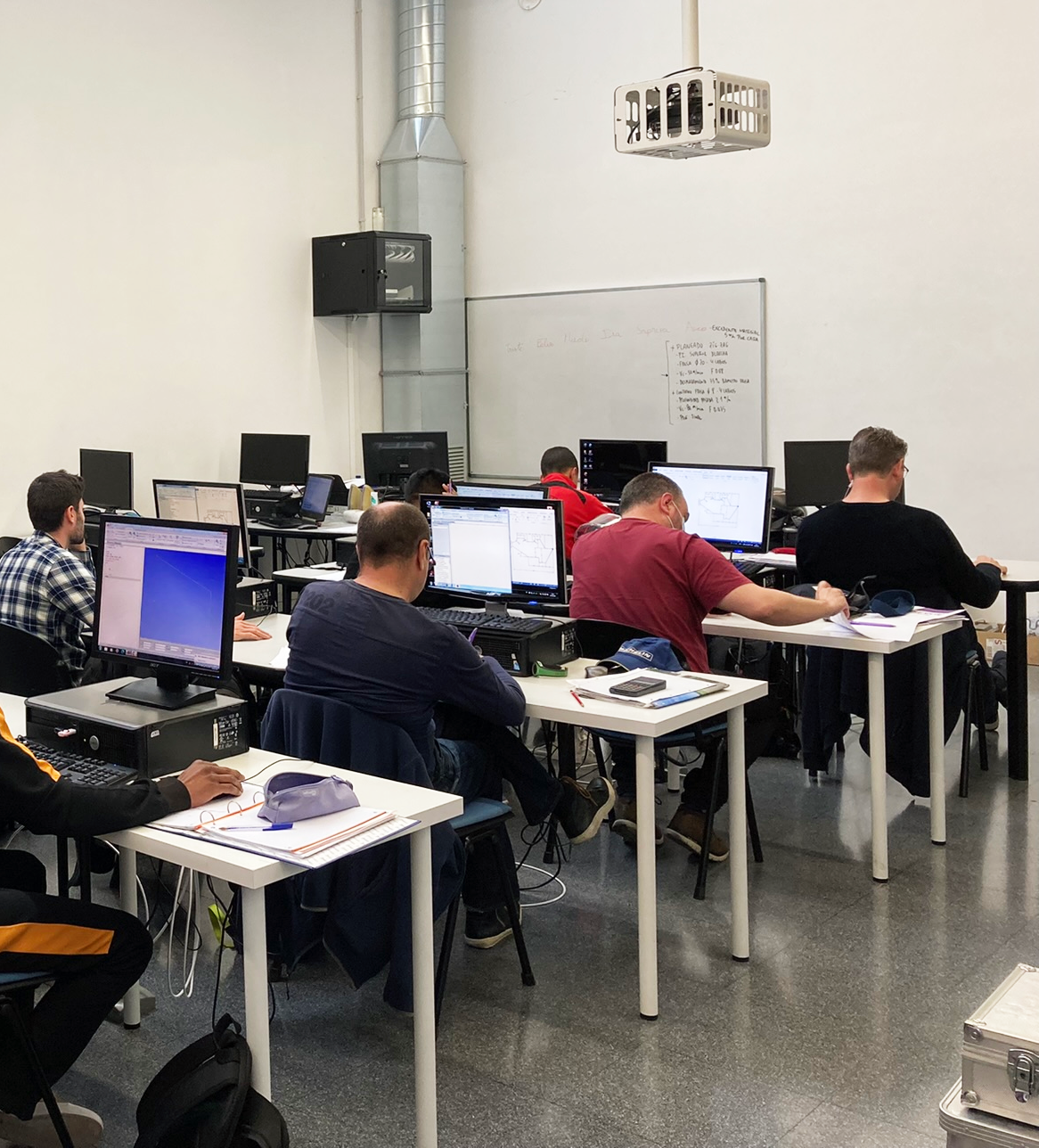}
    \caption{Students taking the exams with the senors wearing.} \label{fig:exam-scenario}
\end{figure}

The following picture (Fig.~\ref{fig:exam-scenario}) is from the class while doing the experiment. 

\begin{figure}[htbp]
    \centering
    \includegraphics[width=0.6\linewidth]{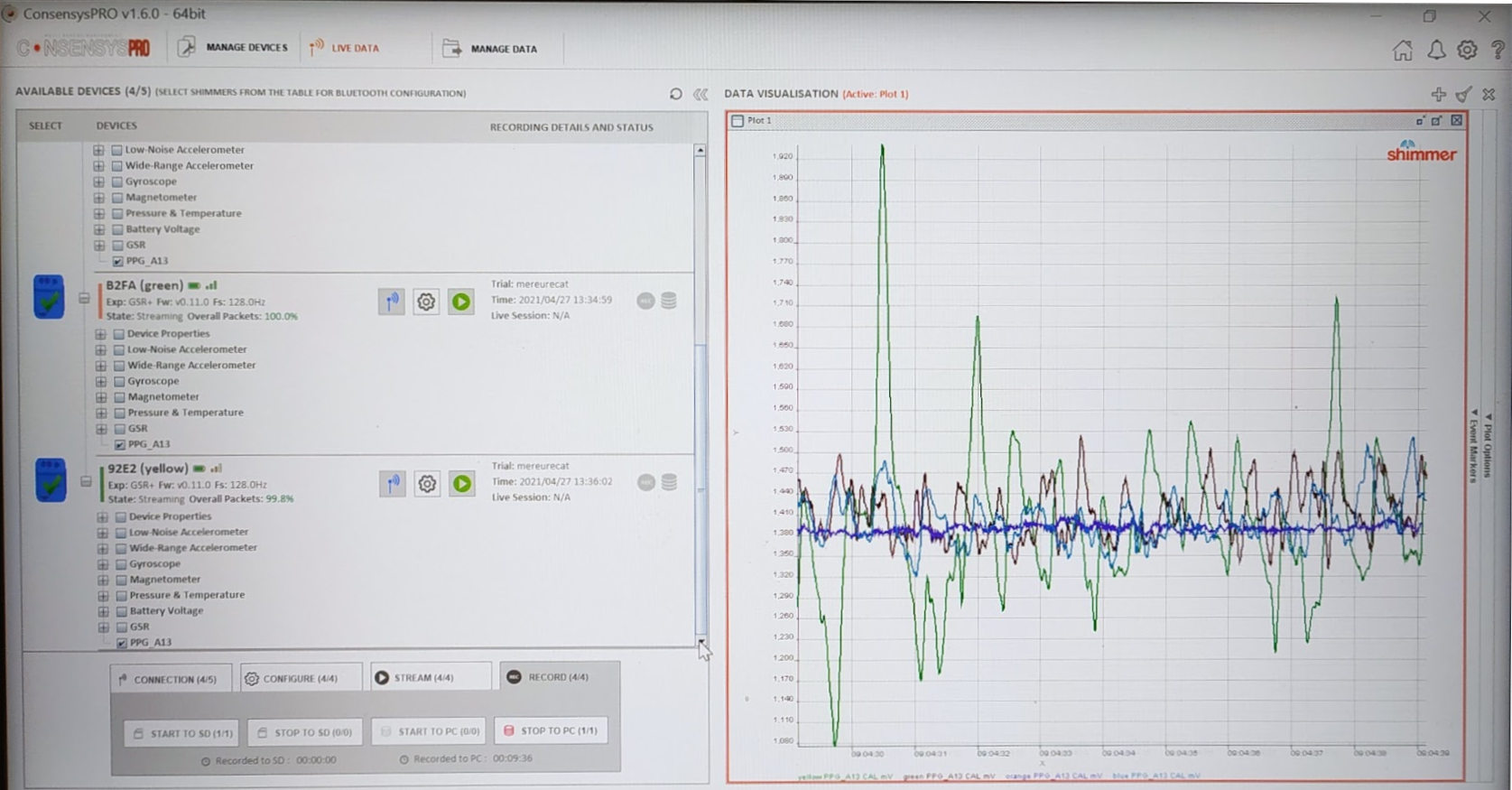}
    \caption{Shimmer's Consensys software interface while collecting the data.}
    \label{fig:software-interface}
\end{figure}

The data collection from multiple Shimmer sensors were done using the Shimmer's Consensys software, in which all the data streams are recorded and saved. In the picture (Fig.~\ref{fig:software-interface}), the software interface while doing the experiment is presented. 

\subsection{Data Formatting}
The data recorded in the Shimmer's Consensys software is exported into two format, such as \textit{.csv} and \textit{.mat}. In each file the data is saved in the following format (Fig.~\ref{fig:file-format}):
\begin{figure}[htbp]
    \centering
    \includegraphics[width=0.5\linewidth]{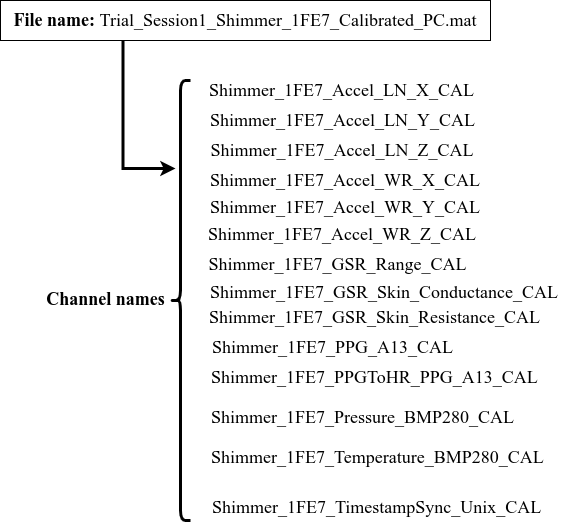}
    \caption{Default file format in Shimmer's Consensys software.}
    \label{fig:file-format}
\end{figure}

Each \textit{.mat} or \textit{.csv} file contains these channels which are available in the Shimmer sensor. From the filename shown in Fig.~\ref{fig:file-format}, we can see that Consensys software saves the file in the following format: \[ [Trial\; name/number, Session number, Sensor Unique ID, Calibrated/not calibrated, PC] \]

\begin{figure}[ht]
    \centering
    \includegraphics[width=0.75\linewidth]{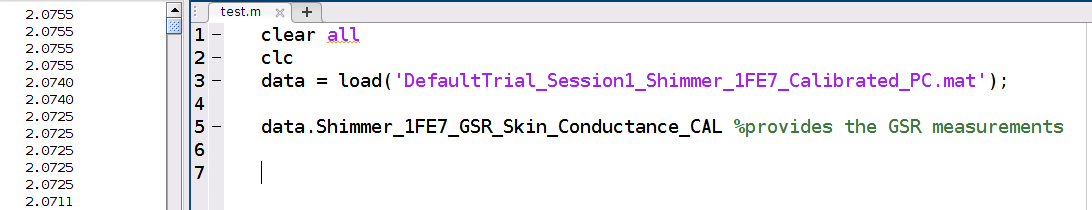}
    \caption{Reading data from the \textit{.mat} file.}
    \label{fig:data-read}
\end{figure}

The sensor recordings can be accessed by providing the channel names as dot(.) operator, e.g. in Fig.~\ref{fig:data-read}.

The emotion labels were saved in .csv file with the file name which contains [session name + student ID + type of data (EDA/PPG)]. 

\subsection{Data Set}

\begin{figure}[ht]
    \centering
    \includegraphics[width=0.5\linewidth]{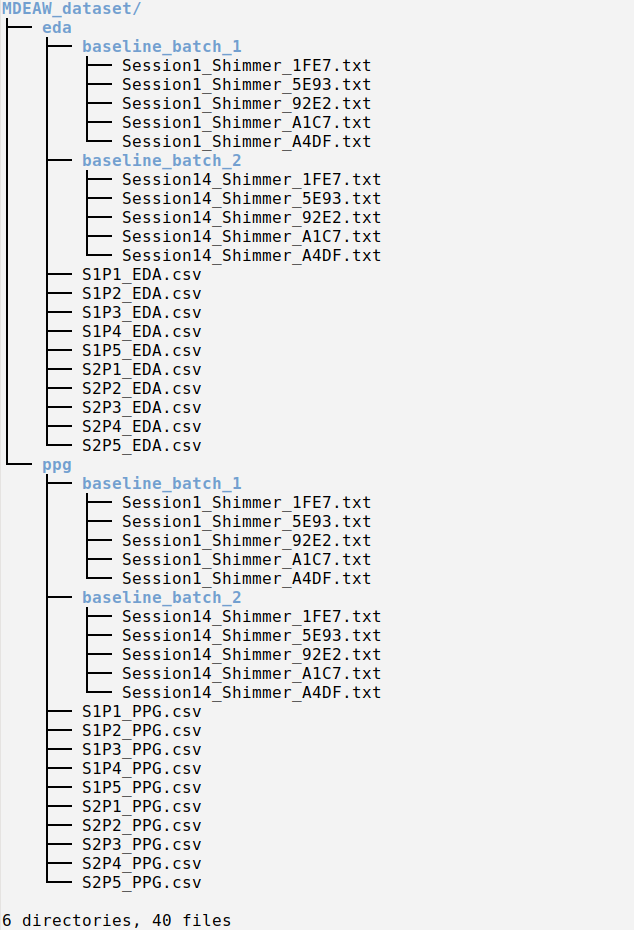}
    \caption{MDEAW dataset folder structure and files inside it.}
    \label{fig:MDEAW-folder-structure}
\end{figure}

For easy accessibility and better understandability we have extracted only the EDA and PPG data from the data recordings saved in Shimmer's Consensys software. The final datatset with the EDA and PPG recordings with the emotion labels are saved in .csv file. The folder structure of the MDEAW dataset is shown in Fig.~\ref{fig:MDEAW-folder-structure}.

Inside each csv file the EDA and PPG recordings are organised as follows:
\[ [ Question no, Session No, Student ID, EDA/PPG data, Emotion labels ] \]

\noindent The MDEAW dataset will be made publicly available.

\begin{figure}[htbp]
    \centering
    \includegraphics[width=0.8\linewidth]{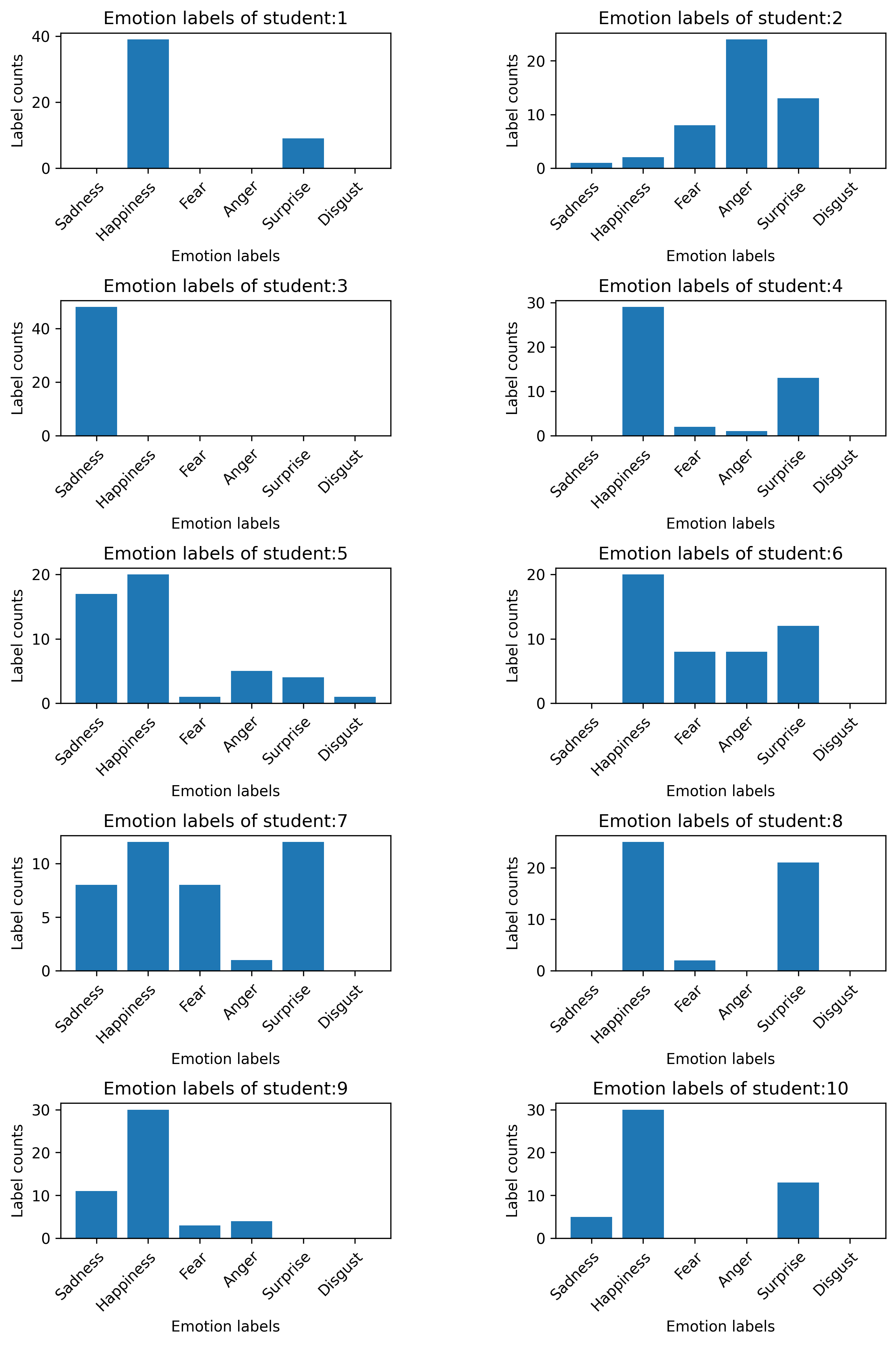}
    \caption{Number of emotion class labels for individual student in MDEAW dataset.}
    \label{fig:individual-student-emotion-class-labels}
\end{figure}

\begin{figure}
    \centering
    \includegraphics[width=0.8\linewidth]{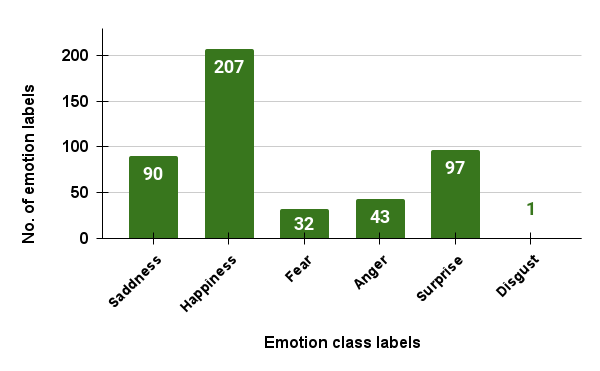}
    \caption{Emotion class distribution in MDEAW dataset}
    \label{fig:emotion-class-all}
\end{figure}

\noindent In Fig~\ref{fig:individual-student-emotion-class-labels}, the emotion class frequencies per student is presented and in Fig.~\ref{fig:emotion-class-all} the overall emotion class frequencies for the MDEAW dataset is presented. This figures shows the class distributions of the MDEAW dataset. 

\section{Analysis of the Data Set}
\subsection{Feature Extraction}
It's important to extract features from different physiological signals (such as EDA, PPG etc.) for retrieving relevant information, which effectively represents emotion states. Based on the extracted features the emotion classifier is tested and trained.

The most popular and widely used time-frequency analysis of various signals (especially EDA, PPG, etc.) for feature extraction is Wavelet Decomposition (WD)~\cite{islam2019}. Its popularity and wide use are due to its localized analysis approach(i,e. time-frequency), multi-rate filtering, and multi-scale zooming. It is better suited for non-stationary signals (such as EDA, PPG etc.)~\cite{ZHANG2020}. Most frequently used wavelet base functions are Meyer WD, Morlet Mother WD, Haar Mother WD and Daubechies WD~\cite{SUBASI2007}. The most frequently used features extracted from each sub-bands of EDA and PPG are entropy, median, mean, standard deviation, variance, 5th percentile value, 25th percentile value, 75th percentile value, 95th percentile value,  root means square value, zero crossing rate, mean crossing rate~\cite{bota2020,Ayata2020,Ayata2016}. 

In our experiment we have extracted and used these features for emotion classification. The wavelet feature extraction technique is used to extract features from multi-modal signal streams (EDA, and PPG signals from the dataset) in this experiment. The wavelet Daubechies 4 (Db4) is the base function for feature extraction. Our experiment decomposes EDA and RB, into three levels, respectively.

\subsubsection{Fusion of EDA and PPG-based features}
The use of features based on multiple modalities has been shown to provide increased classification accuracy compared to approaches based on a single modality. In order to evaluate the performance of the combined EDA and PPG-based features, the two feature vectors $F_{EDA}$ and $F_{PPG}$ are fused as follows: First, the values of each feature vector are normalised in the range [0, 1] in order to compensate for the differences in numerical range. Then, the two normalised feature vectors $F_{EDA}$ and $F_{PPG}$ are concatenated in the final feature vector $F_{fused}$ = [ $F_{EDA}$ $F_{PPG}$].

\section{Emotion Recognition Results for the Data Set}

For the emotion recognition results, we have used our previously developed Real-time Multimodal Emotion Classification System (ReMECS~\cite{remecs}) based on Feed-Forward Neural Network, trained in an online fashion using the Incremental Stochastic Gradient Descent algorithm. The avg. accuracy and F1-score along with the standard deviation (STD) are presented in TABLE~\ref{tab:result}. More results can be found in Appendix~\ref{app:further-results}.

\begin{table}[htbp]
\centering
\caption{Avg. accuracy and F1-score of real-time emotion classification from MDEAW multimodal dataset}\label{tab:result}
\begin{tabular}{|c|c|c|}
\hline
\textbf{\begin{tabular}[c]{@{}c@{}}Dataset\\ Name\end{tabular}} & \textbf{\begin{tabular}[c]{@{}c@{}}Avg. \\ Accuracy\end{tabular}} & \textbf{\begin{tabular}[c]{@{}c@{}}Avg.\\ F1-score\end{tabular}} \\ \hline
\textbf{MDEAW} & 0.9583 ($\pm$ 0.1998) & 0.9583 ($\pm$ 0.1998) \\\hline
\end{tabular}
\end{table}

\begin{figure}[htbp]
    \centering
    \includegraphics[width=0.5\linewidth]{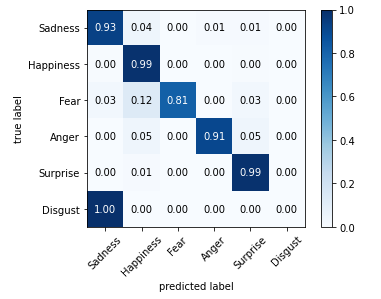}
    \caption{Confusion matrix showing the individual class accuracy for MDEAW using ReMECS}
    \label{fig:confusion-matrix}
\end{figure}

Also, the confusion matrix is presented in Fig.~\ref{fig:confusion-matrix}. 

\section{A New Round of the Experiment}

We have planned a new round of the experiment, to extend the dataset. The reason for conducting another round of experiment to make the MDEAW dataset more diverse and also to increase number data samples by increasing number participants. The diversity can help each procedure to guarantee a total good machine learning: diversity of the training data ensures that the training data can provide more discriminative information for the model, diversity of the learned model (diversity in parameters of each model or diversity among different base models) makes each parameter/model capture unique or complement information and the diversity in inference can provide multiple choices each of which corresponds to a specific plausible local optimal result~\cite{diversity-ml-dl}.  Having more number of diverse data samples in the MDEAW dataset will help ML/DL models to generalize more so that the accuracy and efficiency of those intelligent emotion models increase.

In the new round of experiment we will follow the same design pattern followed for the first experiment for the EDA and PPG data collection from the students but the course and the subject may be different or same based on the availability. The reason for adding the term "based on availability" is because we do the experiment on real students with the real courses studied at our Eurecat Academy.

Right now, the dataset has 10 students data, which is rather small and if we train (offline mode training) any ML/DL model, there might be a chance that the model will be over-fitted easily. So, by increasing the data samples in the MDEAW dataset we can avoid the issue so that ML/DL models can learn and generalize properly and those trained models can be used for emotion classification from new EDA and PPG signals. Also, making the MDEAW dataset diverse, balanced so that the ML/DL model's accuracy can be good for emotion classification.

\section*{Acknowledgement}
\noindent The authors would like to thank the tutors and students at Eurecat - Centro Tecnológico de Cataluña for their participation and collaboration in the experimental study.

This study is carried out as part of the Ph.D work of the student Arijit Nandi and has been funded by ACCIÓ, Catalunya, Spain (Pla d’Actuació de Centres Tecnològics 2021) under the project TutorIA for three years, from December 2019 to December 2022.

\section*{Disclaimer}
\noindent While every care has been taken to ensure the accuracy of the data included in the MDEAW dataset,the authors and the Eurecat - Centro Tecnológico de Cataluña, Barcelona, Spain do not provide any guaranties and disclaim all responsibility and all liability (including without limitation, liability in negligence) for all expenses, losses, damages (including indirect or consequential damage) and costs which you might incur as a result of the provided data being inaccurate or incomplete in any way and for any reason 2022, Eurecat - Centro Tecnológico de Cataluña, Barcelona, Spain.

\section*{Contact}
\noindent For any questions regarding the MDEAW database please contact: mail@eurecat.org
Eurecat - Centro Tecnológico de Cataluña, Barcelona, Spain.

\bibliographystyle{IEEEtran}
\bibliography{reference.bib}

\begin{thebibliography}{10}
\providecommand{\url}[1]{#1}
\csname url@samestyle\endcsname
\providecommand{\newblock}{\relax}
\providecommand{\bibinfo}[2]{#2}
\providecommand{\BIBentrySTDinterwordspacing}{\spaceskip=0pt\relax}
\providecommand{\BIBentryALTinterwordstretchfactor}{4}
\providecommand{\BIBentryALTinterwordspacing}{\spaceskip=\fontdimen2\font plus
\BIBentryALTinterwordstretchfactor\fontdimen3\font minus
  \fontdimen4\font\relax}
\providecommand{\BIBforeignlanguage}[2]{{%
\expandafter\ifx\csname l@#1\endcsname\relax
\typeout{** WARNING: IEEEtran.bst: No hyphenation pattern has been}%
\typeout{** loaded for the language `#1'. Using the pattern for}%
\typeout{** the default language instead.}%
\else
\language=\csname l@#1\endcsname
\fi
#2}}
\providecommand{\BIBdecl}{\relax}
\BIBdecl

\bibitem{ekman1992}
\BIBentryALTinterwordspacing
P.~Ekman, ``An argument for basic emotions,'' \emph{Cognition and Emotion},
  vol.~6, no. 3-4, pp. 169--200, 1992. [Online]. Available:
  \url{https://doi.org/10.1080/02699939208411068}
\BIBentrySTDinterwordspacing

\bibitem{islam2019}
M.~R. {Islam} and M.~{Ahmad}, ``Wavelet analysis based classification of
  emotion from eeg signal,'' in \emph{Int'l Conf. on Electrical, Computer and
  Comm. Eng.}, 2019, pp. 1--6.

\bibitem{ZHANG2020}
J.~Zhang, Z.~Yin, P.~Chen, and S.~Nichele, ``Emotion recognition using
  multi-modal data and machine learning techniques: A tutorial and review,''
  \emph{Information Fusion}, vol.~59, pp. 103 -- 126, 2020.

\bibitem{SUBASI2007}
A.~Subasi, ``Eeg signal classification using wavelet feature extraction and a
  mixture of expert model,'' \emph{Expert Systems with Applications}, vol.~32,
  no.~4, pp. 1084 -- 1093, 2007.

\bibitem{bota2020}
P.~Bota, C.~Wang, A.~Fred, and H.~Silva, ``Emotion assessment using feature
  fusion and decision fusion classification based on physiological data: Are we
  there yet?'' \emph{Sensors}, vol.~20, no.~17, 2020.

\bibitem{Ayata2020}
D.~Ayata, Y.~Yaslan, and E.~Kamasak, Mustafa, ``Emotion recognition from
  multimodal physiological signals for emotion aware healthcare systems,''
  \emph{J. of Medical and Biological Eng.}, pp. 149--157, 2020.

\bibitem{Ayata2016}
D.~{Ayata}, Y.~{Yaslan}, and M.~{Kamaşak}, ``Emotion recognition via random
  forest and galvanic skin response: Comparison of time based feature sets,
  window sizes and wavelet approaches,'' in \emph{Medical Technologies National
  Congress}, 2016, pp. 1--4.

\bibitem{remecs}
A.~Nandi, F.~Xhafa, L.~Subirats, and S.~Fort, ``Real-time multimodal emotion
  classification system in e-learning context,'' in \emph{Proceedings of the
  22nd Engineering Applications of Neural Networks Conference}.\hskip 1em plus
  0.5em minus 0.4em\relax Cham: Springer International Publishing, 2021, pp.
  423--435.

\bibitem{diversity-ml-dl}
Z.~Gong, P.~Zhong, and W.~Hu, ``Diversity in machine learning,'' \emph{IEEE
  Access}, vol.~7, pp. 64\,323--64\,350, 2019.

\bibitem{fed-remecs}
\BIBentryALTinterwordspacing
A.~Nandi and F.~Xhafa, ``A federated learning method for real-time emotion
  state classification from multi-modal streaming,'' \emph{Methods}, vol. 204,
  pp. 340--347, 2022. [Online]. Available:
  \url{https://www.sciencedirect.com/science/article/pii/S104620232200072X}
\BIBentrySTDinterwordspacing

\bibitem{fed-remecs-u}
A.~Nandi, F.~Xhafa, L.~Subirats, and S.~Fort, ``Federated learning with
  exponentially weighted moving average for real-time emotion classification,''
  in \emph{Proceedings of the 13th 13th International Symposium on Ambient
  Intelligence Conference}.\hskip 1em plus 0.5em minus 0.4em\relax Cham:
  Springer International Publishing, 2022, pp. 423--435.

\end{thebibliography}

\begin{appendices}
\section{Learning Examination Exercises}\label{app:exam}

\includegraphics[width=\linewidth]{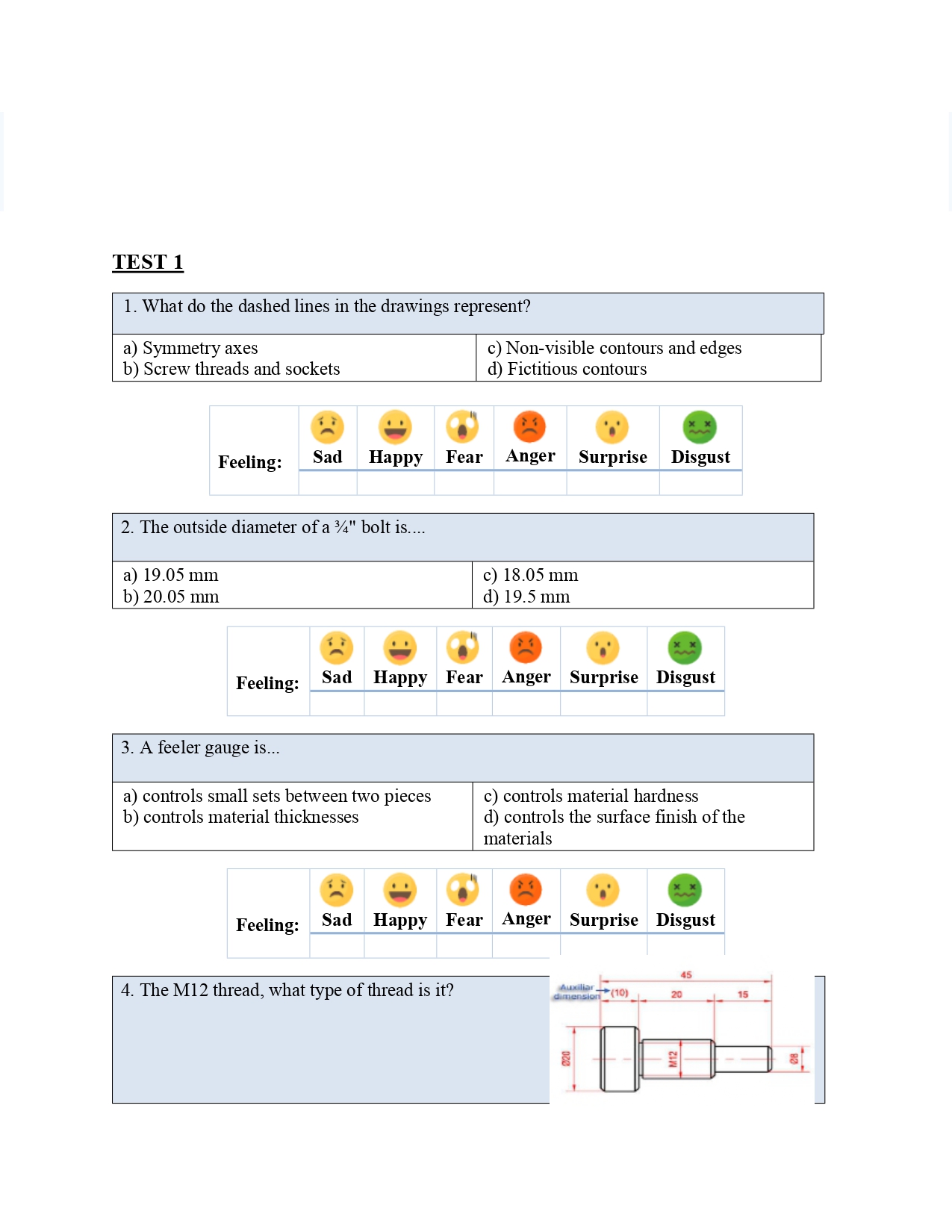}
\includegraphics[width=\linewidth]{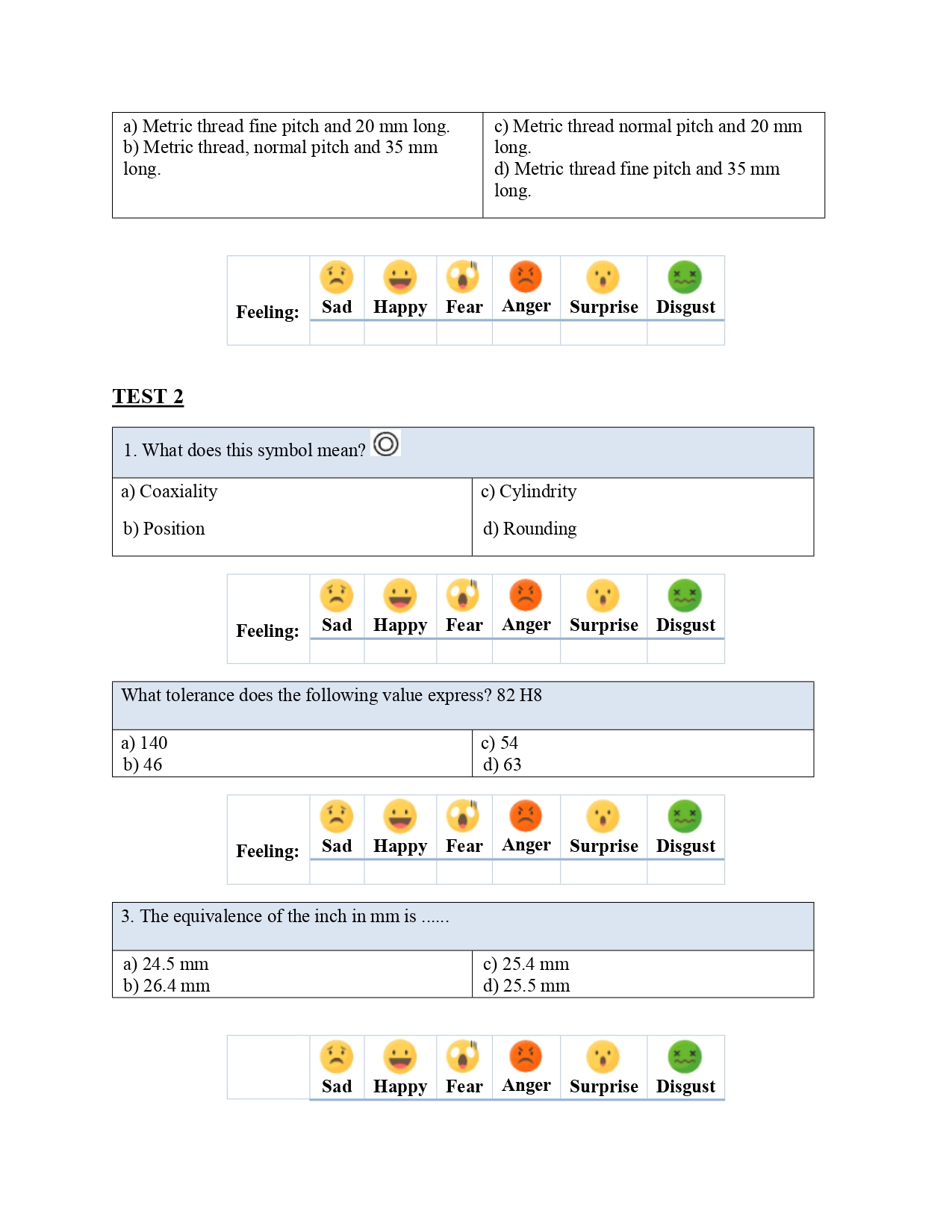}
\includegraphics[width=\linewidth]{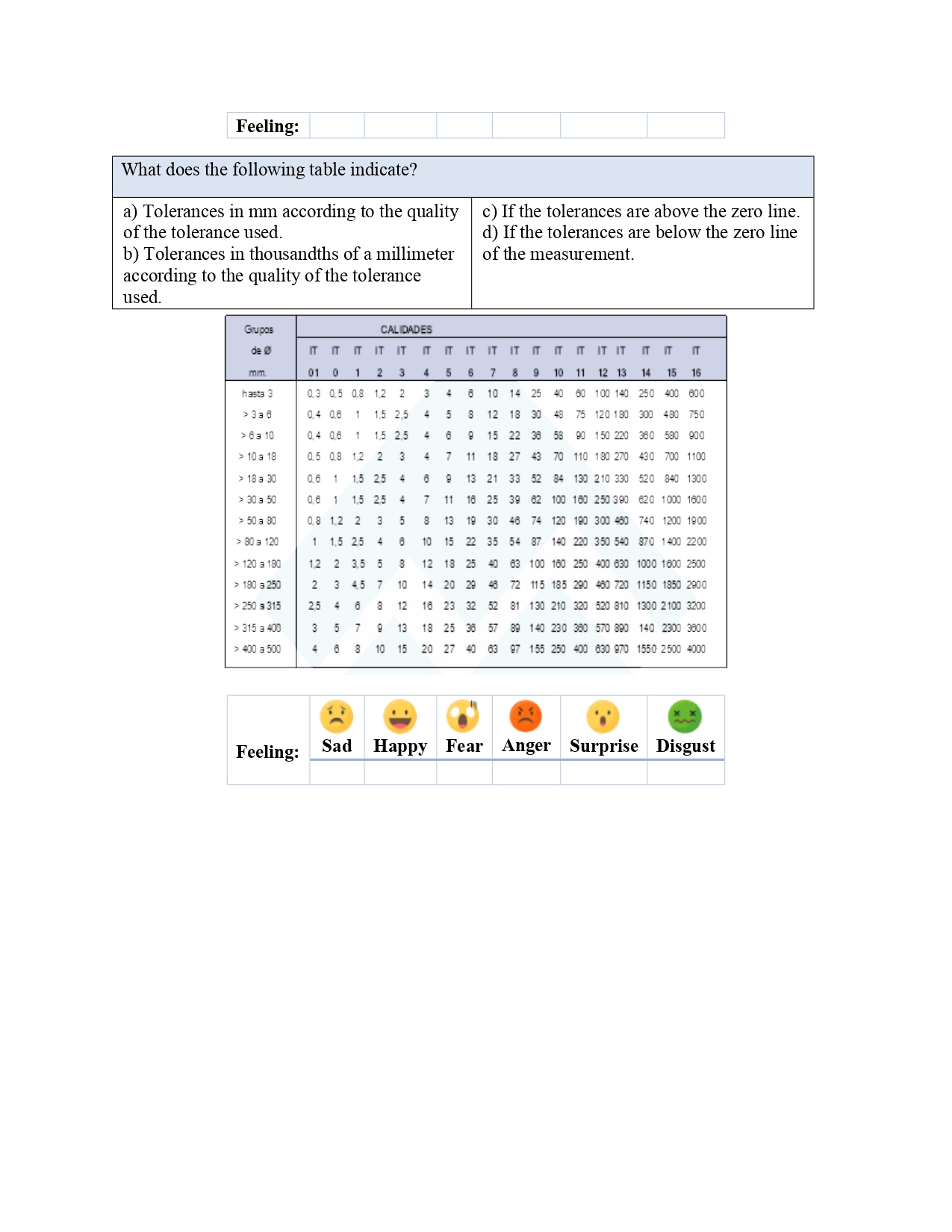}
\includegraphics[width=\linewidth]{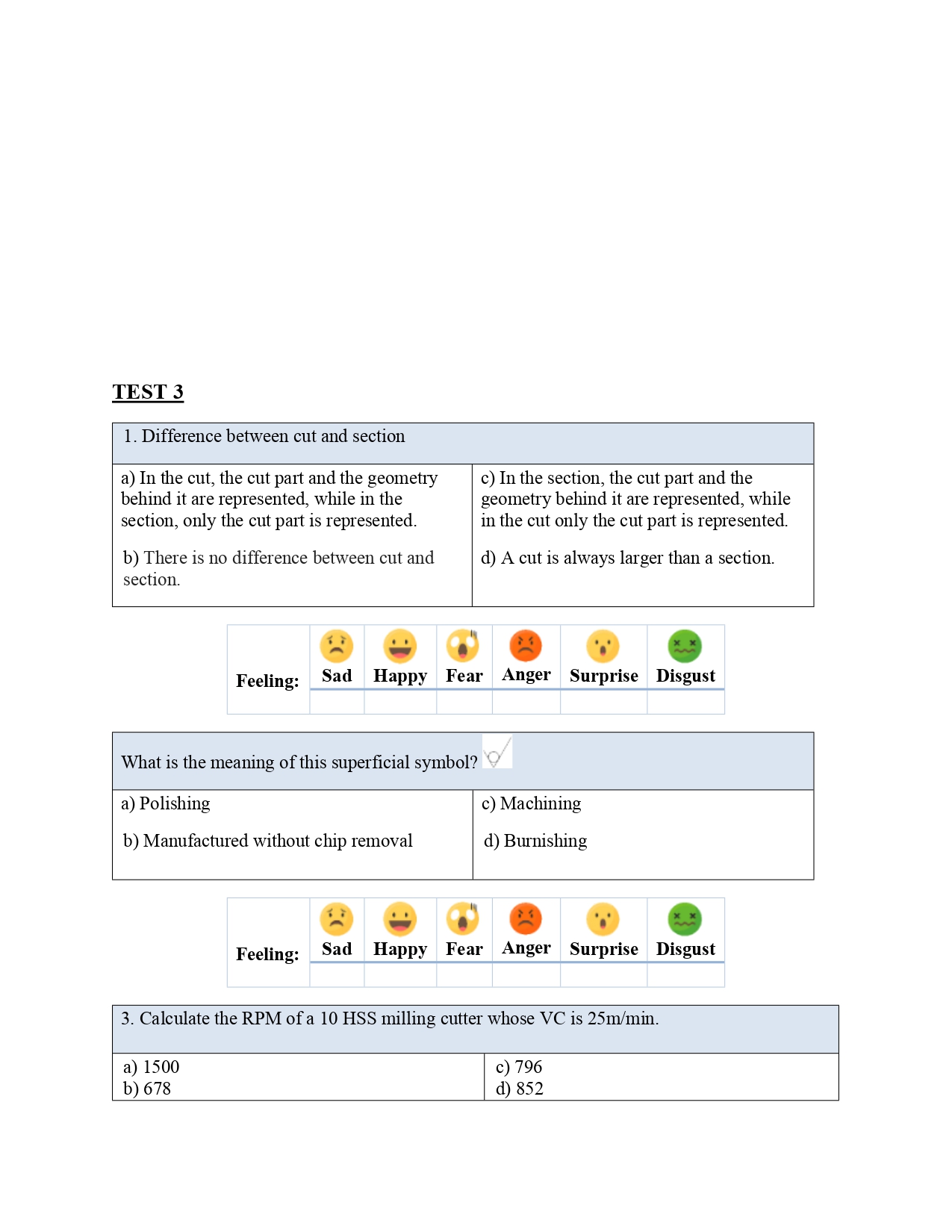}
\includegraphics[width=\linewidth]{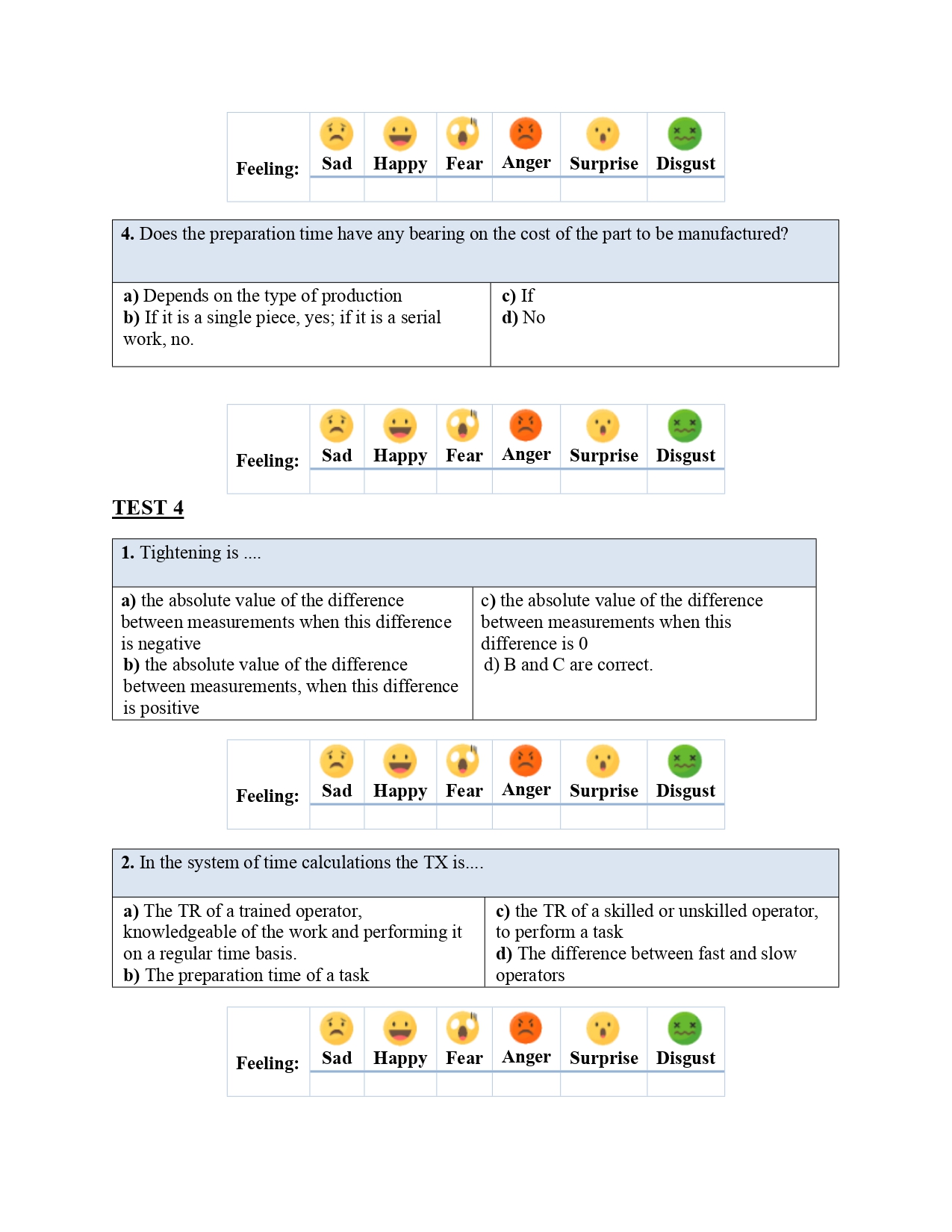}
\includegraphics[width=\linewidth]{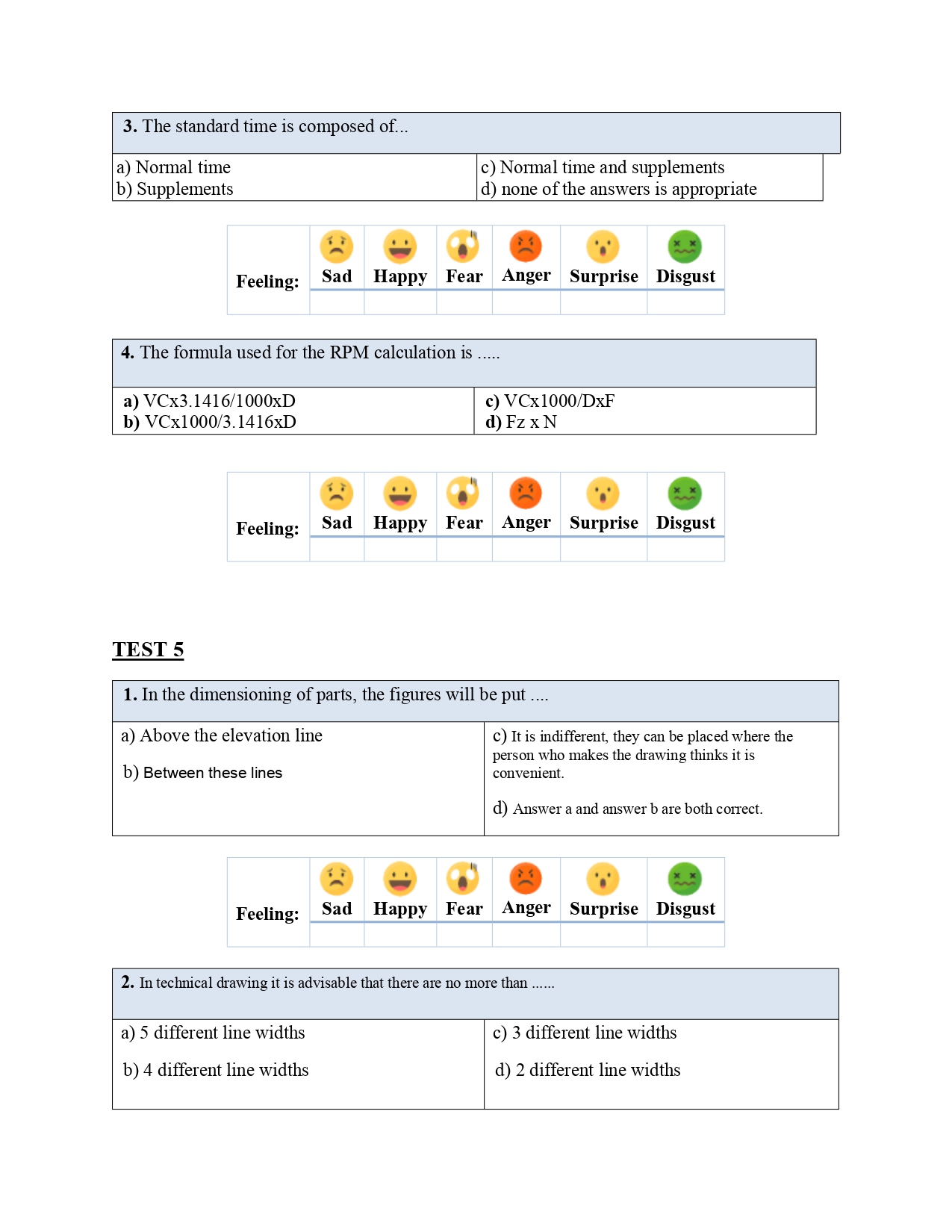}
\includegraphics[width=\linewidth]{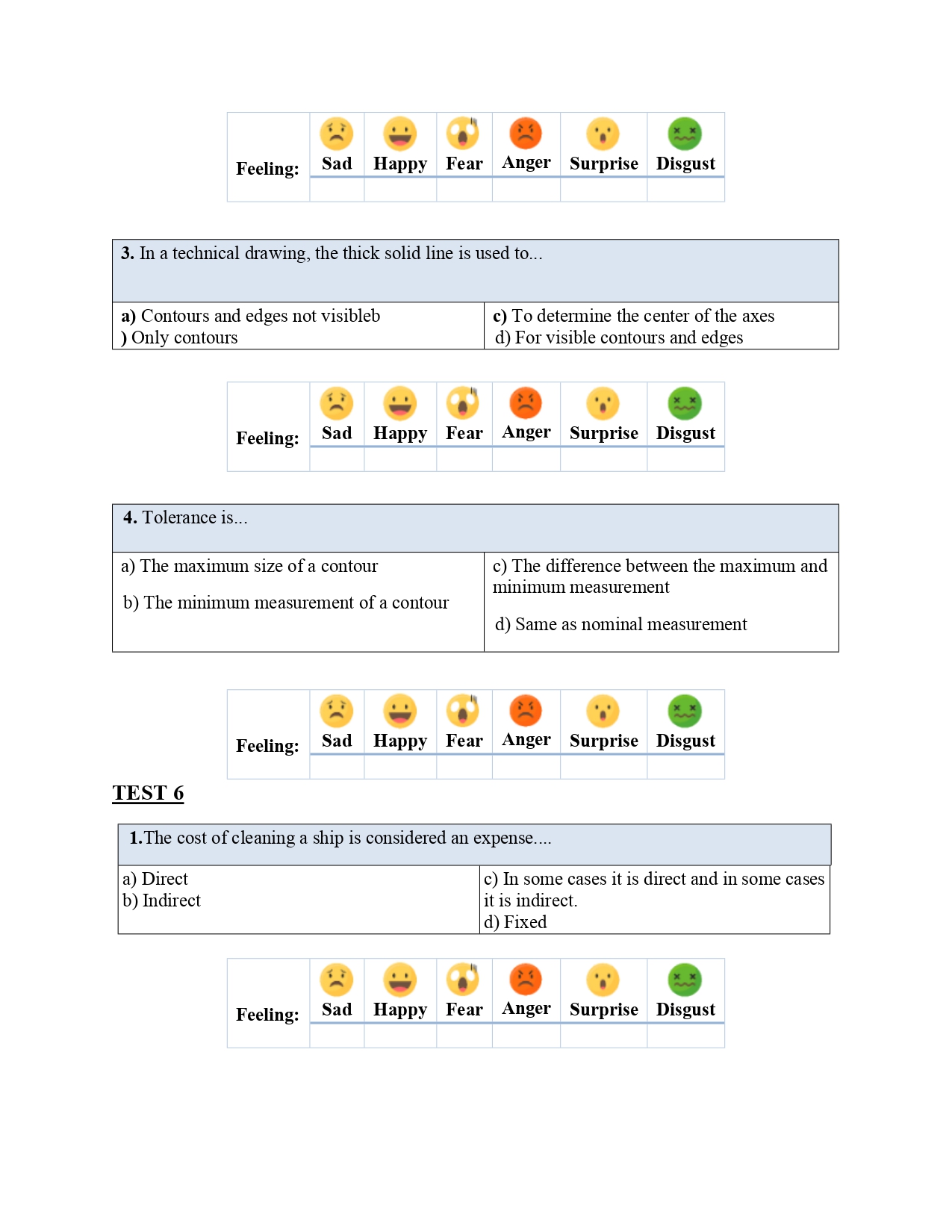}
\includegraphics[width=\linewidth]{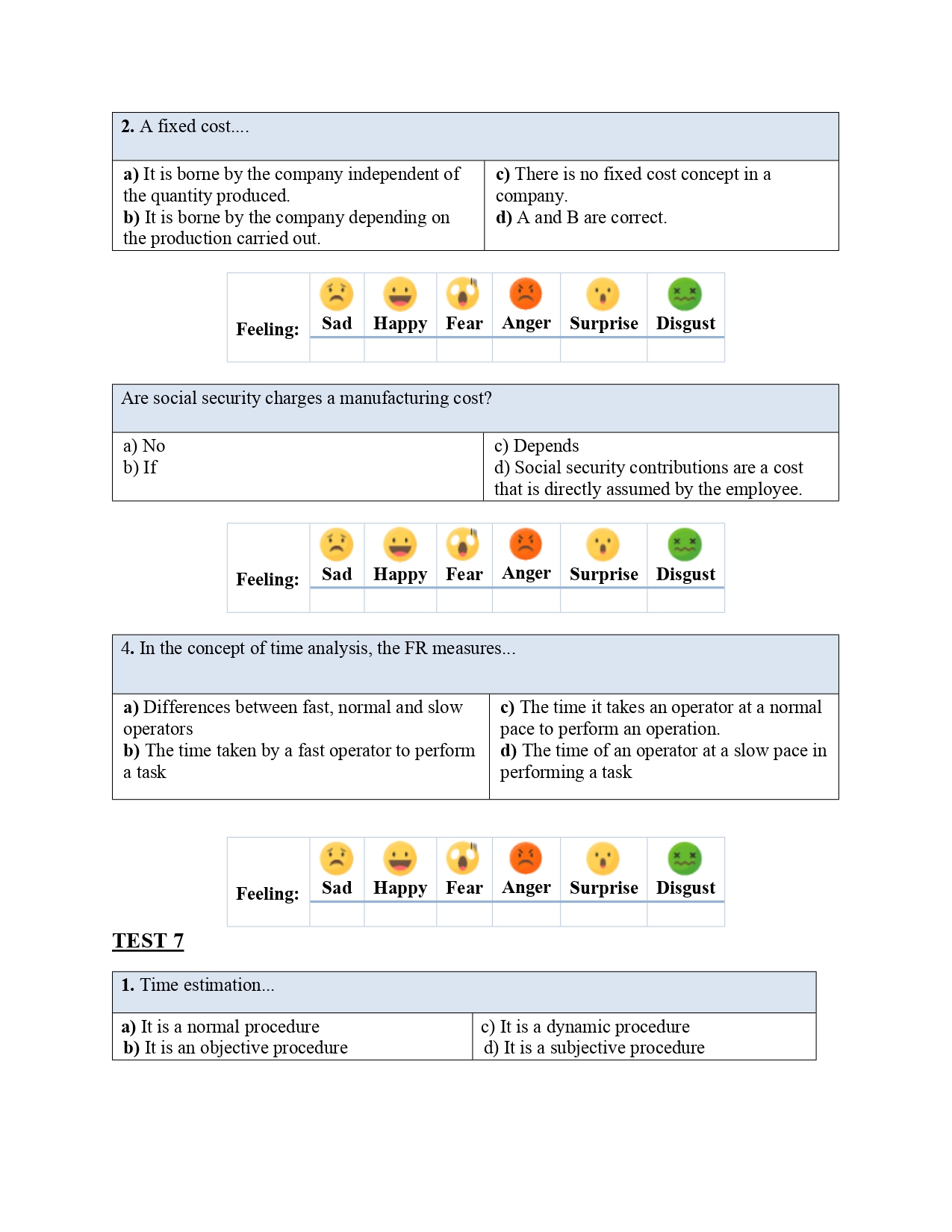}
\includegraphics[width=\linewidth]{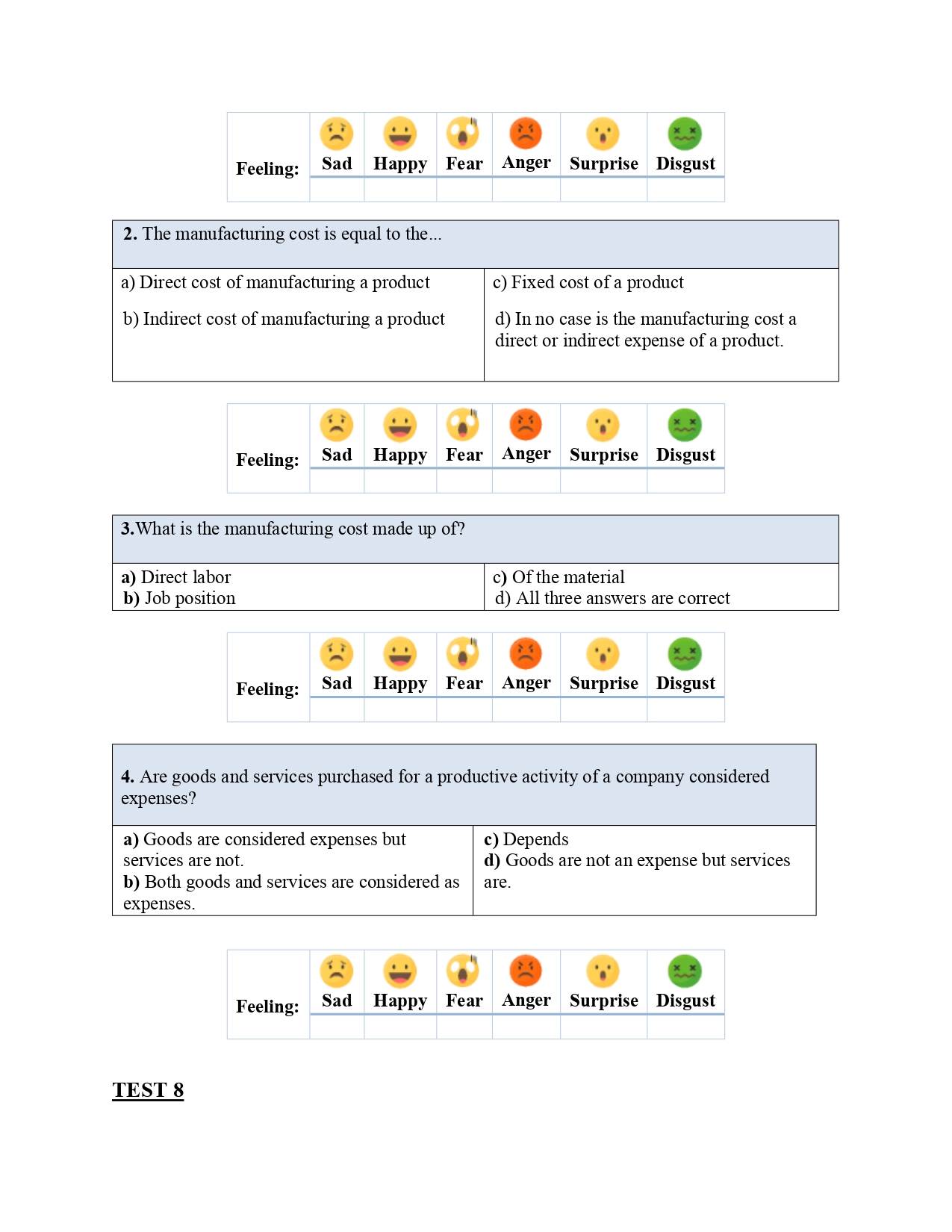}
\includegraphics[width=\linewidth]{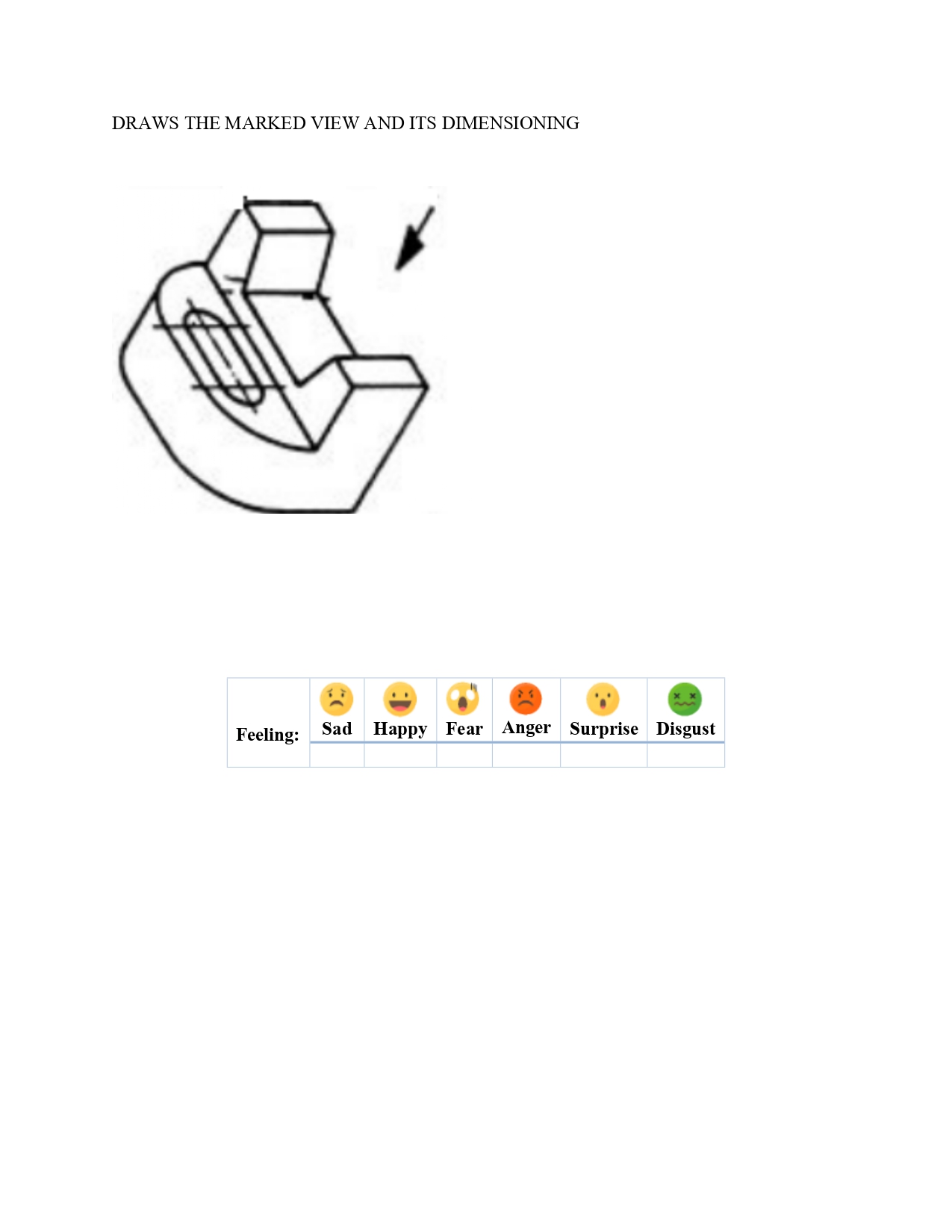}
\includegraphics[width=\linewidth]{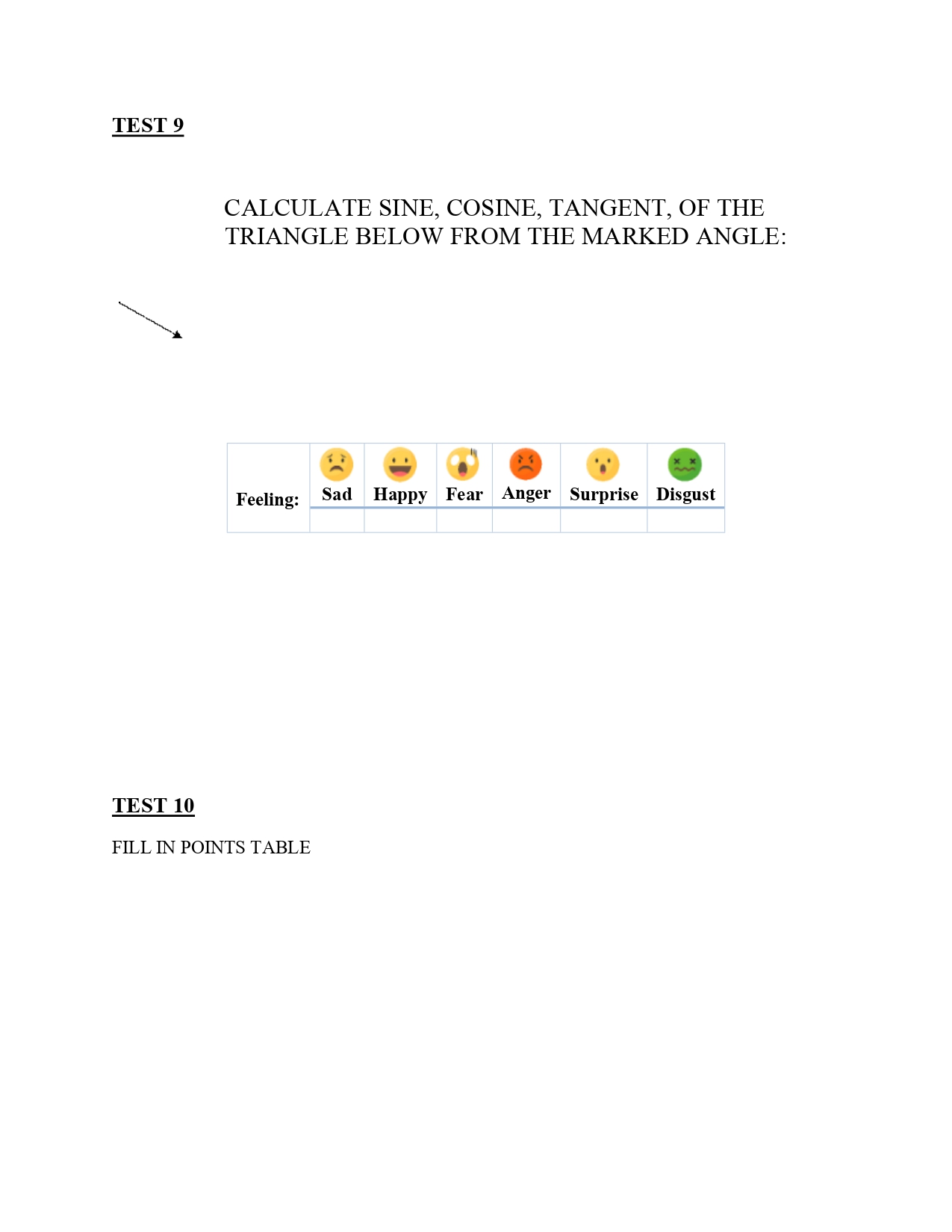}
\includegraphics[width=\linewidth]{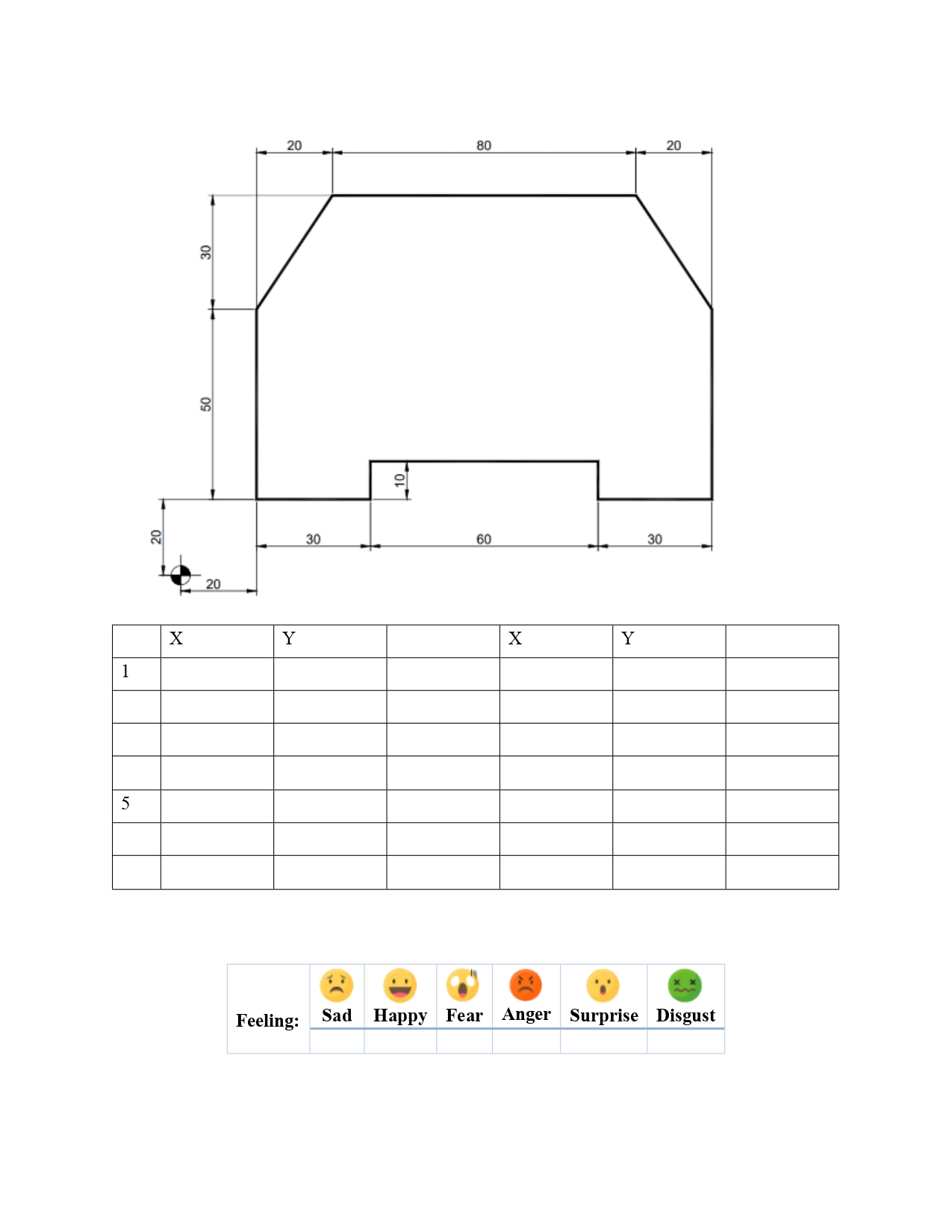}
\includegraphics[width=\linewidth]{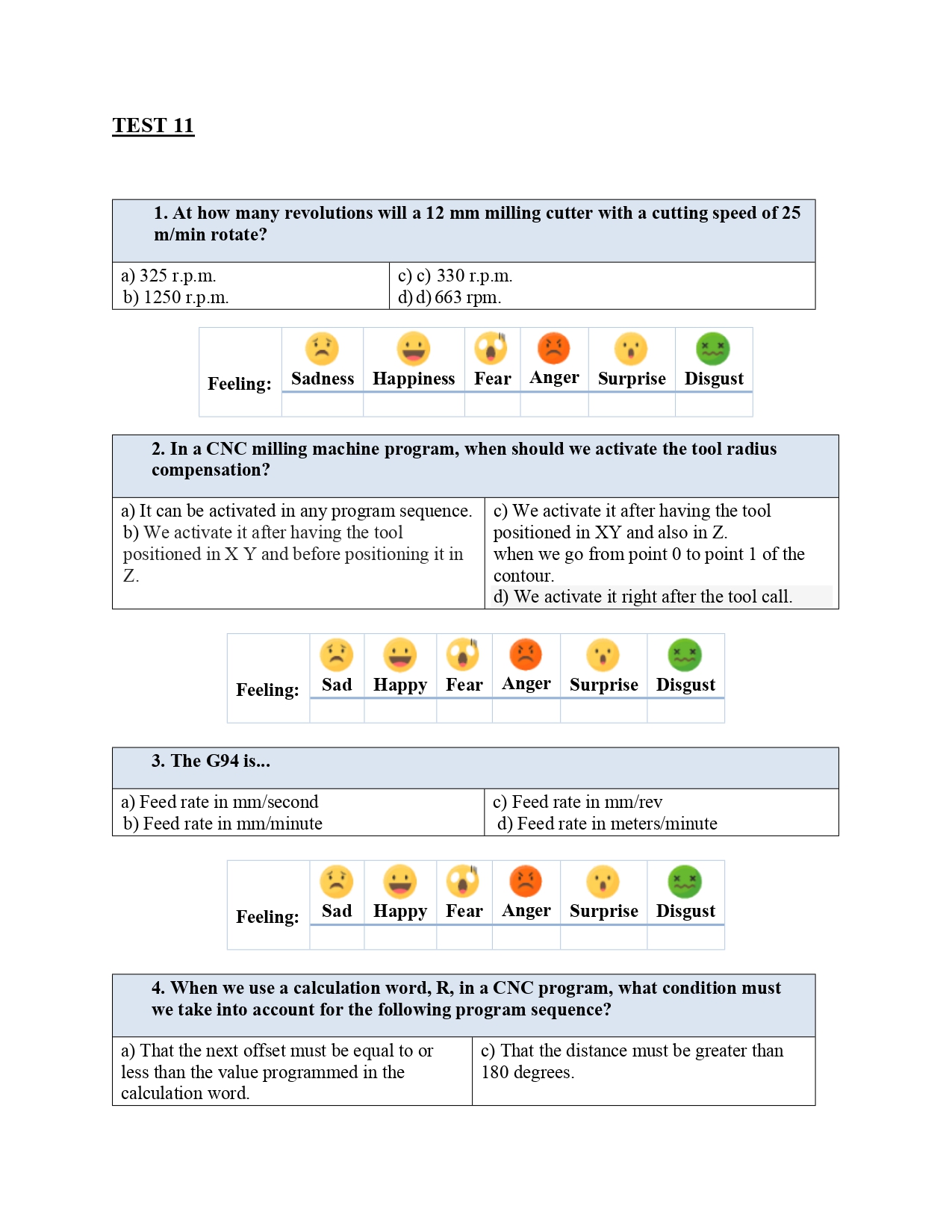}
\includegraphics[width=\linewidth]{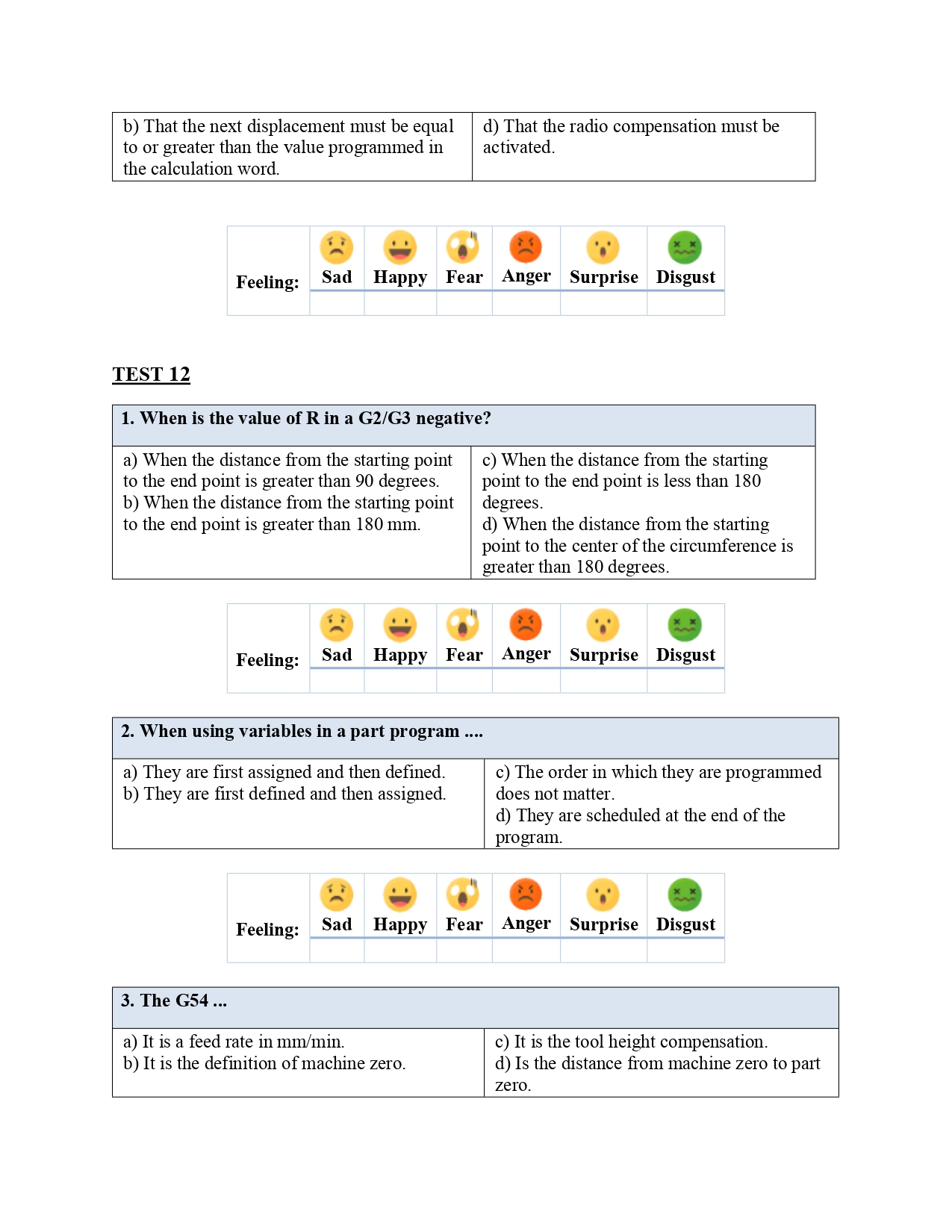}
\includegraphics[width=\linewidth]{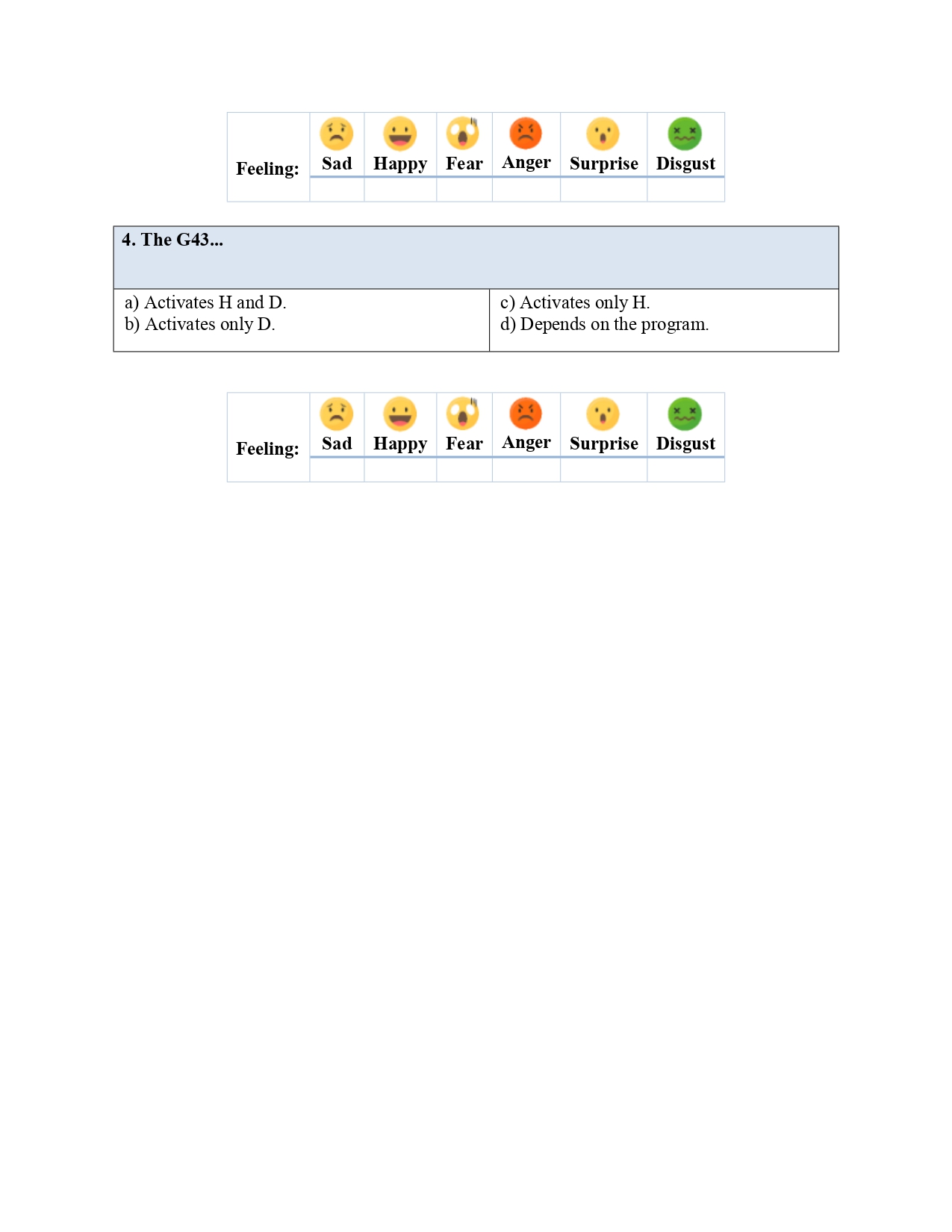}


\section{Further Data Analysis Results}\label{app:further-results}
We have applied our previously developed frameworks of Fed-ReMECS~\cite{fed-remecs}  and Fed-ReMECS-U~\cite{fed-remecs-u} with the MDEAW dataset. Herewith, we show some of the results.

\subsection{Results from Fed-ReMECS Framework}\label{fed-remecs}

\begin{table}[htbp]
\centering
\caption{Global model's avg. accuracy in case of Fed-ReMECS~\cite{fed-remecs}}
\label{tab:result-fed-remecs}
\resizebox{0.45\linewidth}{!}{\begin{tabular}{|c|c|}
\hline
\textbf{No. of clients} & \textbf{Accuracy} \\ \hline
5 & 0.9315 ($\pm$0.16) \\ \hline
10 & 0.6195 ($\pm$0.25) \\ \hline
\end{tabular}}
\end{table}

\begin{figure}[htbp]
    \centering
    \includegraphics[width=0.7\linewidth]{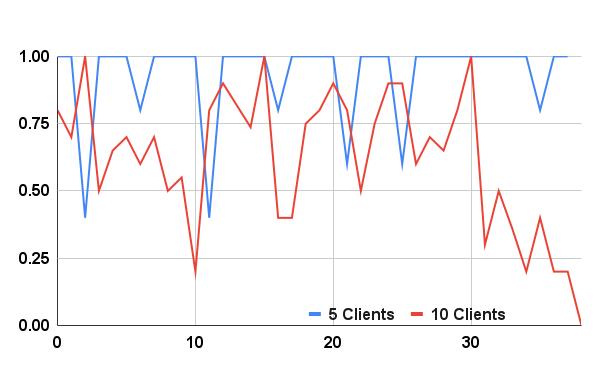}
    \caption{Global model's performance in Fed-ReMECS for different client scenario (5 and 10 clients) using MDEAW dataset.}
    \label{fig:fed-remecs}
\end{figure}

The global model performance in case of Fed-ReMECS is shown in Fig.~\ref{fig:fed-remecs} 

\subsection{Results from Fed-ReMECS-U Framework}\label{fed-remecs-u}:

\begin{table}[ht]
\centering
\caption{Global model's avg. accuracy in case of Fed-ReMECS-U~\cite{fed-remecs-u}}
\label{tab:result-fed-remecs-u}
\resizebox{0.45\linewidth}{!}{\begin{tabular}{|c|c|}
\hline
\textbf{No. of clients} & \textbf{Accuracy} \\ \hline
5 & 0.9487 ($\pm$0.13) \\ \hline
10 & 0.9410 ($\pm$0.15) \\ \hline
\end{tabular}}
\end{table}

\begin{figure}[ht]
    \centering
    \includegraphics[width=0.7\linewidth]{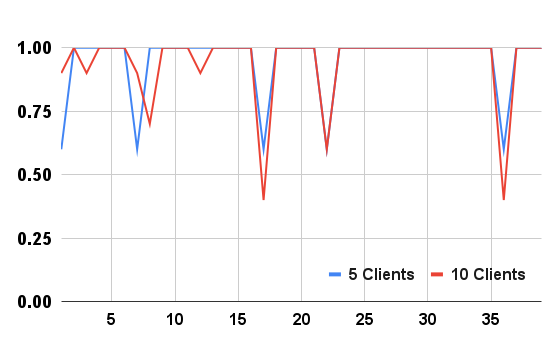}
    \caption{Global model's performance in Fed-ReMECS-U for different client scenario (5 and 10 clients) using MDEAW dataset.}
    \label{fig:fed-remecs-u}
\end{figure}

The global model performance in case of Fed-ReMECS-U is shown in Fig.~\ref{fig:fed-remecs-u} 

\end{appendices}

\end{document}